\documentclass{ws-ijmpd}
\usepackage[super,compress]{cite}
\usepackage{graphicx}
\usepackage[english]{babel}
\usepackage{graphicx}
\usepackage{bm}
\usepackage{amsmath}
\usepackage{amssymb}
\usepackage{amsfonts}
\usepackage{epsfig}
\usepackage{colordvi}
\usepackage{color}
\begin{document}

\markboth{S.  Capozziello, M. De Laurentis, O. Luongo}
{$f(R)$  cosmology}

%
\catchline{}{}{}{}{}
%



\title{CONNECTING EARLY AND LATE UNIVERSE BY  $f(R)$ GRAVITY }


\author{Salvatore Capozziello}
\address{Dipartimento di  Fisica, Universit\`a di Napoli ``Federico II", Via Cinthia, I-80126, Napoli, Italy\\
Istituto Nazionale di Fisica Nucleare (INFN), Sez. di Napoli, Via Cinthia, I-80126 Napoli, Italy.\\
Gran Sasso Science Institute (INFN), Viale F. Crispi, 7, I-67100, L'Aquila, Italy.
capozzie@na.infn.it}

\author{Mariafelicia De Laurentis}
\address{Tomsk State Pedagogical University, Tomsk, ul. Kievskaya 60, 634061 Russian Federation\\
mfdelaurentis@tspu.edu.ru}

\author{Orlando Luongo}
\address{Dipartimento di  Fisica, Universit\`a di Napoli ``Federico II", Via Cinthia, I-80126, Napoli, Italy\\
Istituto Nazionale di Fisica Nucleare (INFN), Sez. di Napoli, Via Cinthia, I-80126 Napoli, Italy.\\
Instituto de Ciencias Nucleares, Universidad Nacional Aut\'onoma de M\'exico (UNAM), M\'exico, DF 04510, Mexico.\\
luongo@na.infn.it}

\maketitle

\begin{history}
\received{Day Month Year}
\revised{Day Month Year}
\end{history}

\begin{abstract}
Inflation and dark energy are two of the most relevant aspects of modern cosmology. These  different epochs provide    the universe is passing through accelerated phases soon after  the Big-Bang  and at present stage of its evolution. In this review paper, we discuss that both  eras can be, in principle, described by a  geometric picture, under the standard of $f(R)$ gravity.
We give the fundamental physics motivations and outline the main ingredients of $f(R)$ inflation, quintessence and cosmography. This wants to be a quick summary of $f(R)$ paradigm without claiming of completeness.
\end{abstract}

\keywords{Alternative Theories of Gravity; Quantum Cosmology; Dark Energy;  Observational Cosmology.}

\ccode{PACS numbers: 04.50.Kd; 98.80.Qc; 95.36.+x; 98.80.-k}

\tableofcontents

\section{Introduction}
\label{zero}

The huge amount of present cosmological data has led to new perspectives and scenarios in the field of modern cosmology \cite{sne1,sne2,sne3,sne4,sne5,sne6}. For the first time, one refers to current-time cosmology as \emph{Precision Cosmology}, {\it i.e.} the cosmological models precisely reproduce the universe expansion history, showing robust bounds which well match cosmic data \cite{copp,beyondcopp}. Relevant consequences of using cosmic data to constrain the correct cosmological models were carried forward from the end of last century. Indeed, before 1998 cosmologists assumed that the total content of the cosmic energy budget was filled by standard pressureless matter density. However, after 1998, several evidences pointed out that the universe is currently undergoing an accelerated expansion. Soon, it was evident that this experimental outcome could not be interpreted by using baryons and dark matter only and so the corresponding standard cosmological model was definitively modified, re-including a cosmological constant term, $\Lambda$, within Einstein's energy momentum tensor \cite{starx,starx2}. The cosmological constant likely represents a first explanation of current universe speeding up \cite{reviy,rax1,costante1,costante2}.
 
The physical nature of  cosmological constant can be related to the existence of non-zero vacuum energy and can be computed in the context of quantum field theory. Unfortunately, theoretical predictions and cosmological observations differ from a huge amount of orders of magnitude, leading to a severe \emph{fine-tuning problem} \cite{weifine}. Moreover, matter density  and $\Lambda$  density are extremely close to each other in order of magnitudes, leading to a further issue named the \emph{coincidence problem} \cite{coincidence}. It consists in the  fact that  there is no reasons to expect that matter and $\Lambda$ densities have to be comparable at present time, since matter evolves as the universe expands, while $\Lambda$ is constant at all stages of the universe evolution. Conversely, to differently assess the observed acceleration,  one may assume that the fluid responsible for the speeding up of the universe cannot be a pure constant along the universe expansion history \cite{padrev,A1,A2,A3,A4,A5,stv1,stv2,stv3,qlcs1,qlcs3,qlcs4,qlcs5,qlcs6}.

In turn, any possible extensions lie on the existence of some additional fluids, whose physical properties, {\it e.g.} particle masses, thermal and electromagnetic interactions, and so forth, are not known \emph{a priori}. Consequently, we do not have a final experimental evidence for the existence of those fluids at a fundamental level. Thus, cosmologists interpreted such a dynamical fluid in terms of a \emph{dark energy} counterpart \cite{darkenergy,darkenergyy,C1}. Afterwards, the need of comparing both late and early-phases of the universe evolution has brought one to wonder whether introducing scalar fields, capable of describing de-Sitter-like phases, would be useful to relate both inflation and current acceleration to a single unified description \cite{campiscalari,campiscalarii,campiscalariii,campiscalariiii}. Hence, alternative approaches have been proposed in terms of curvature invariants and geometric corrections \cite{raxx,lobo1,lobo2,lobo3}. Those schemes, supported by evidences at ultra violet scales, usually involve the use of additional curvature terms into the Hilbert-Einstein action. This recipe enables one to assume that "scalar fields" are derived from geometrical properties of space-time and also provides viable interpretations of dark energy and inflation as geometric effects at large scales (weak energies) and small scales (high energies), respectively \cite{revvia}.

In other words, this geometric view represents a way to generalize and extend the standard General Relativity,  aimed to consistently describe the early-time inflation and late-time acceleration \cite{revvia2,D1,D2}, without  introducing  other \emph{by hand dark components} \cite{PRsergei,PhysRepnostro,defelice}. More practically, a set of extended  theories of gravity,  containing additional  curvature terms, can be relevant at very high energies, naturally producing inflation \cite{staro}. During the cosmic evolution, the curvature decreases and General Relativity gives a sufficiently good  approximation at  intermediate scales. Afterwards, infra-red corrections start to work at very large scales. Rephrasing it differently, the curvature decreasing permits sub-dominant terms to start growing and then transition from deceleration to acceleration to  happen. An important consequence is that this phenomenon roughly fixes the critical points of the whole cosmic evolution \cite{faro}. Thus, the early-time as well as the late-time cosmic speed-up can be  addressed
by the fact that some curvature corrections to the Ricci scalar $R$ provide significative consequences at large and small curvatures \cite{againRmore,bamba1,bamba2,bamba3,bamba4,alter1,aur1,aur2,aur3}.
In summary, a Lagrangian like
\begin{equation}
f(R)\simeq ...+\alpha_{(-2)}R^{-2}+ \alpha_{(-1)}{R^{-1}}+\alpha_{(0)} R^{0} + \alpha_{(1)} R+ \alpha_{(2)} R^2 +...
\end{equation}
or, in general,
\begin{equation}
\label{serie}
f(R) \simeq \sum_{i=-n}^{i=n} \alpha_{(i)}R^i\,,
\end{equation}
with $n \in \mathbb{N}$,  could grossly fit the whole universe expansion history starting from the high energy regimes, $(n > 0)$, recovering the cosmological constant, $(n=0)$, and General Relativity $(n=1)$ at intermediate scales,  and evolving towards infra-red limit at  large scales ($n <0$).  Clearly, the role played by $\Lambda \equiv \alpha_{(0)} R^{0}$ is only formally equivalent to the case of a pure cosmological constant model. Indeed, according to  $f(R)$ gravity there is no  \emph{a priori} reason to consider the cosmological constant as associated to the vacuum energy density. In other words, one may recover a dynamical effective $\Lambda$, assuming it as a limiting case of a more general solution of the above $f(R)$ expansion series\cite{grek,grekbis}. In  doing so, the dynamics of dark energy is not mimicked by $\Lambda$, which appears only as a  zero order term of $f(R)$ gravity.
This may be clearer if one assumes the effective  gravitational action coming from  some fundamental theory. In fact, we do not need to add the cosmological constant as a further term, put by hand into the gravitational action, but to reproduce it from first principles, with a completely different physical interpretation.  In fact, it can be derived for $f(R)$ gravity,  a class of models capable of  producing viable cosmology (different from the $\Lambda$CDM) where the cosmological constant is zero in flat space-time, but appears  in a curved one for sufficiently large curvatures. A {\it smoking gun} for these models could be  the slope of  primordial perturbation power spectrum determined from   CMB fluctuations\cite{raxx}. 

On the other hand,  Lagrangians such as Eq. (\ref{serie}) show several defects and issues that need to be necessarily addressed (see \cite{PRsergei} for a detailed discussion). Unfortunately, this class of drawbacks remains one of the main open problems of modern high-energy physics. However, in the absence of a final quantum gravity theory,  modified gravities can be viewed as practicable approaches built up to comply observational data with  space-time phenomenology \cite{sotfar,farro1,farro2}. In addition, actually as a by-product of this framework,  modified gravities even provide a self-consistent dark matter explanation \cite{lobo0}. It is also possible to  describe, in fact, the observed evidence at galactic and extragalactic scales of dark matter distribution in terms of geometric modifications \cite{annalen}. In this review, we underline how inflation and dark energy can be encompassed within a single geometric approach, offered by $f(R)$ gravity which can be considered as  the simplest geometrical extension of General Relativity. We summarize, with no claims of completeness, the most relevant clues related to $f(R)$ theories and  get hints on possible future developments.  This approach does not exhaust the possibilities of extended theories where more general curvature invariants can be employed \cite{PhysRepnostro}, albeit $f(R)$ models may be assumed as a useful paradigm \cite{sotfar2}.

The paper is structured as follows. Sec. \ref{due}  is  a quick  summary on the emergence of curvature corrections as soon as a quantum field theory is formulated on curved space.  In  Sec. \ref{tre}, we consider a realization of such an approach:  the Starobinsky model capable of naturally producing an   inflationary scenario. In Sec. \ref{quattro}, we develop the variational principles and the field equations of $f(R)$ gravity. The basic equations of $f(R)$ cosmology are presented in Sec. \ref{cinque} where some toy models, in view of dark energy, are discussed. Sec. \ref{sei} is a wide discussion of cosmography where cosmographic parameters are constructed starting from $f(R)$ functions and their derivatives. The goal is to recover viable phenomenological models  in agreement with observational  data. Conclusions are drawn in Sec.\ref{sette}.

\section{Curvature corrections  from fundamental physics}
\label{due}

Let us start our discussion  with some fundamental physics considerations.
At high energies and small scales, an accurate description of matter requires quantum field theory formulated on
  curved spaces \cite{BirrellDavies}.
Since the matter should be quantized, one
can assume a semi-classical description of gravitation where
Einstein's equations take the form
 \begin{eqnarray}
G_{\mu\nu}  \equiv R_{\mu\nu} -\frac{1}{2} \, g_{\mu\nu} R =<T_{\mu\nu}> \,.
 \label{Intro1.2.1}
\end{eqnarray}
Here, we are far from the full quantum gravity regime and $<T_{\mu\nu}>$ is the expectation value of the quantum stress-energy tensor acting as a  source in  the gravitational field. Hereafter, we assume physical units where conventionally $8\pi G=k_B=c=1$.  The l.h.s of the above field equations is assumed to classically evolve. The simplest case is the homogeneous and isotropic universe, characterized by a Friedmann-Robertson-Walker metric
 \begin{eqnarray}
 \label{FRW}
ds^2=dt^2-a^2(t)d\Omega_{ k}^2\,,
\end{eqnarray}
where $d\Omega_k^2$ is the metric on the 3-space whose topology depends on the space-curvature parameter $k$.
In a curved space-time, also in the case in which both matter and radiation fluids are zero, quantum fluctuations of  fields determine non-trivial contributions to the whole energy-momentum tensor
\cite{BirrellDavies,parkerbook} and may arise. In the presence of conformal invariant, massless and free-matter fields, those corrections can be framed as:
 \begin{eqnarray}
<T_{\mu\nu}> = k_{1}\,  ^{(1)}H_{\mu\nu} + k_{3}\,^{(3)}H_{\mu\nu} \,,
 \label{Intro1.2.5}
\end{eqnarray}
with $k_1$ and $k_3$ numerical coefficients and
\begin{eqnarray}
^{(1)}H_{\mu\nu} & = &   2R_{;\mu\nu}-2g_{\mu\nu} \Box R +2RR_{\mu\nu}
-\frac{1}{2} \, g_{\mu\nu} R^{2}\,, \label{Intro1.2.6} \\
&&\nonumber\\
 ^{(3)}H_{\mu\nu} & = &  {R^{\sigma}}_{\mu} R_{\nu\sigma}
-\frac{2}{3} \, R R_{\mu\nu} -\frac{1}{2} \, g_{\mu\nu}
 R^{\sigma\tau}R_{\sigma\tau} +\frac{1}{4} \, g_{\mu\nu} R^{2}\,.
 \label{Intro1.2.7}
\end{eqnarray}
An important remark is useful at this point. The masses of the matter fields and their mutual interactions  
can be neglected in the high curvature limit because $R>>
m^{2}$. The matter-graviton interactions generate   
non-minimal 
coupling terms in the effective 
Lagrangian. The 
one-loop contributions of such terms are comparable to the ones 
coming from (\ref{Intro1.2.5}) and generate, from the conformal point 
of view, the same effects on gravity. The simplest effective 
Lagrangian that takes into account these corrections is
\begin{eqnarray}
{\cal L}_{NMC} =- \frac{1}{2} \nabla^{\alpha} \phi 
\nabla_{\alpha} \phi -V(\phi)  - \frac{\xi}{2} \,  
R\phi^{2} \,, \label{Intro1.2.23}
 \end{eqnarray}
where $\xi$ is a dimensionless coupling constant between 
the scalar and the gravitational fields. The scalar field 
stress-energy  tensor  will be modified accordingly but  a conformal 
transformation can be found 
such that the modifications due to curvature terms can, 
at least formally, be cast in the form of a matter-curvature 
interaction\cite{PhysRepnostro}. The same argument holds for the trace anomaly, as we will see below.

\noindent The tensor $^{(1)}H_{\mu\nu}$ is conserved, since $^{(1)}H_{\mu;\nu}^{\nu}=0$. This tensor is obtained by varying a quadratic
contribution of the Ricci scalar $R$ in the local action,
\begin{eqnarray}
^{(1)}H_{\mu\nu}
=\frac{2}{\sqrt{-g}} \frac{\delta}{\delta g^{\mu\nu}}
\left( \sqrt{-g}\,
R^{2} \right) \,.    \label{Intro1.2.9}
\end{eqnarray}
The infinities coming from  $ <T_{\mu\nu}>$ should be neglected somehow. To do so and to obtain a re-normalized theory, one might add an infinite number of many counterterms in the Lagrangian density of  gravity. One of those terms is, for example, $C R^{2}\sqrt{-g}$, where $C$ represents a diverging parameter in terms of a logarithm. In addition, one has to consider
\begin{eqnarray}
 ^{(2)}H_{\mu\nu} = 2{R^{\sigma}}_{\mu ; \nu \sigma}-\Box
R_{\mu\nu}- \frac{1}{2} \, g_{\mu\nu}\Box R +{R^{\sigma}}_{\mu} R_{\sigma
\nu}-\frac{1}{2} \, R^{\sigma\tau} R_{\sigma\tau}g_{\mu\nu}\,,
\end{eqnarray}
where the relation
\begin{eqnarray}
 ^{(2)}H_{\mu\nu} = \frac{1}{3} \, ^{(1)}H_{\mu\nu}\,
 \end{eqnarray}
holds in conformally flat space-times. In these cases, only the
first and the third  $H_{\mu\nu} $ terms of Eq. (\ref{Intro1.2.5}) do not vanish.
Since  one can add to  the  term
$C \sqrt{-q} \, R^2  $ an arbitrary constant, the
coefficient $k_{1}$ may assume any value and,  in principle, should be
determined experimentally \cite{BirrellDavies,parkerbook}. On the other hand, the tensor $ ^{(3)}H_{\mu\nu} $ is conserved only in conformally
flat space-time and it
cannot be obtained by varying a local action. Finally, one has \cite{BirrellDavies}
\begin{eqnarray}
k_{3}=\frac{1}{1440\pi^{2}} \left( N_{0}+\frac{11}{2} \, N_{1/2}
+31 N_{1}\right)\,,   \label{Intro1.2.10}
 \end{eqnarray}
where the  coefficients $N_i$'s ($i=0, 1/2, 1$) are given by the number of  quantum  fields with spin $0,1/2$, and $1$ present into the dynamics. Moreover, vector fields would give their contributions highly to $k_3$ due to the larger coefficient $ 31 $ showed in $N_1$. The cited  massless fields,
as well as the spinorial case, are even described by conformally invariant equations. They are present in $ <T_{\mu\nu}>$ in the  form
(\ref{Intro1.2.5}). The energy-momentum tensor trace goes to zero for conformally invariant classical fields whereas, owing to the term weighted by $k_{3}$, one infers that the outcome derived from the  tensor (\ref{Intro1.2.5}) provides a non-vanishing trace. This leads to the existence of the  {\it trace anomaly} which may show serious consequences in cosmology. The matter field masses and the corresponding mutual interactions may be neglected as $R>>m^{2}$, {\it i.e.} at high curvature regime as discussed above. In addition,  interactions between matter and gravitons lead to non-minimally coupled terms in the effective field Lagrangian. Summing up, what we have found definitively forecast that, as one quantizes matter fields on curved space-times, higher-order curvature corrections naturally arise as a corresponding effect. The paradigm deals with the fact that generic higher-order curvature corrections to the Hilbert-Einstein action can be easily framed at fundamental levels and the corresponding effects are highly relevant  both at ultra-violet and at infra-red energy scales.

\section{The case of Starobinsky inflation}
\label{tre}

A realization of the above effective theory is the Starobinsky inflation \cite{staro}, where higher curvature terms give the possibility to realize a de-Sitter behavior for the early universe.
The Starobinsky model represents a prototype of any $f(R)$ cosmology that, in principle, can track the whole cosmic history as soon as the cosmological solutions fit dynamics of the various epochs (e.g. transit from accelerated to decelerated behaviors and viceversa \cite{PRsergei}).  Let us define the quantities \cite{vilenkin}:
\begin{eqnarray}
\label{2.7}
H_0=(k_3 )^{-\frac{1}{2}}\,,\qquad M=(6k_1 )^{-\frac{1}{2}}\,,
\end{eqnarray}
which have the obvious meaning of the Hubble parameter $H_0$ related to the number of quantum fields  and an effective mass $M$.
The above tensor, defined in Eq. (\ref{Intro1.2.5}), can be re-written as
\begin{eqnarray}
\label{2.8}
 <T_{\mu\nu}> =\frac{1}{H_0^{2}}\,^{(3)}H_{\mu\nu}+\frac{1}{6M^{2}}\, ^{(1)}H_{\mu\nu}\,,
\end{eqnarray}
where, for physical compatibility, we place $H_0>0$ and $M>0$.
Even though the trace of the energy-momentum tensor is null for conformally invariant fields, the expected value of Eq. (\ref{2.8}) has a non-zero trace,  that is
\begin{eqnarray}
\label{2.9}
 <T_{\nu}^{\nu}> =\frac{1}{H_0^{2}}\left( \frac{1}{3}R^2-R_{\nu\sigma}R^{\nu\sigma}\right)-\frac{1}{M^{2}}R_{;\nu}\,^{;\nu}\,.
\end{eqnarray}
This  trace anomaly means that the conformal invariance
is broken by the regularization of infinities in the energy-momentum tensor \cite{BirrellDavies}.
Eq. (\ref{Intro1.2.1}), with $<T_{\mu\nu}> $ given by Eq.
(\ref{2.8}),  contains  a de-Sitter space-time
\begin{eqnarray}
\label{2.10}
R_{\mu\nu}=\frac{1}{4}\,g_{\mu\nu}\,R\,,\quad R=\mbox{const}\,,
\end{eqnarray}
as a possible solution.
Substituting Eq. (\ref{2.10}) into  Eqs. (\ref{Intro1.2.1}) and (\ref{2.8}) and discarding  the
trivial solution $R =0$, we obtain $R =12H_0$. The corresponding
de-Sitter solutions are
\begin{eqnarray}
\label{2.11}
a (t) &=&H_0^{-1}\cosh(H_0\,t)\,, \quad k =+1\,,\nonumber\\
a (t) &=&a_0 \exp(H_0\,t)\,, \quad k =0\,,\\
a (t) &=&H_0^{-1}\sinh (H_0\,t)\,, \quad k=-1\,,\nonumber
\end{eqnarray}
for closed, flat, and open models, respectively. These
solutions describe  inflationary phases driven  by quantum curvature corrections of Einstein's equations.
 The  $H_0$ value depends on the numbers of fields
involved in  Eq. (\ref{Intro1.2.10}).  Typically, it is not so much different from the
Planck mass $m_p$.   For example, in the minimal $SU(5)$ model, it is 
$N_0=34$, $N_{1/2}= 45$, $N_1=24$, $8\pi k_{3}=1.8$, and then $H_0=0.7m_p$. Clearly the value of $H_0$ evolves according to the number of quantum fields present into dynamics. In particular, it changes after the inflation and for  any phase transition.
The evolution equation for the scale factor, obtained
from Eqs. (\ref{Intro1.2.1}) and (\ref{2.8}), by inserting the Friedmann-Robertson-Walker metric (\ref{FRW}), is \cite{staro,vilenkin}:
\begin{eqnarray}
\label{2.12}
\frac{{\dot a}^2+k}{a^2}=\frac{1}{H^2_0}\left[\frac{{\dot a}^2+k}{a^2}\right]^2-\frac{1}{M^2}\left[2 \frac{{\dot a}{\dddot a}}{a^2}-\frac{{\ddot a}^2}{a^2}+2\frac{{\ddot a}{\dot a}^2}{a^3}-3\left(\frac{\dot a}{a}\right)^4-2k\frac{{\dot a}^2}{a^4}+\frac{k^2}{a^4}\right]\,.\nonumber\\
\end{eqnarray}
It is worth noticing that the source of the Friedmann equation, (i.e. the r.h.s), is totally geometric.  In Eq. (\ref{2.12}), we indicate with $k$ the spatial curvature scalar, {\it i.e.} the curvature of the spatial part of Einstein's equations. In $\Lambda$CDM cosmology and in inflation,  the scalar curvature $k$ is negligibly small and it is usually neglected, albeit it is not completely clear if its role may influence the dark energy evolution\cite{sgh}.

The de-Sitter solutions (\ref{2.11}) implies that the universe scale factor exponentially grows and the  $k$-dependent terms in
(\ref{2.12}) become negligible. It is, therefore, sufficient to
study the fiat space model with $k=0$.
Introducing  $H(t)={\dot a}/a$, we can rewrite
Eq. (\ref{2.12})  as

\begin{eqnarray}
\label{2.13}
H^2\left(H^2-H_0^2\right)=\frac{H_0^2}{M^2}\left(2H{\ddot H}+6H^2{\dot H}-{\dot H}^2\right).
\end{eqnarray}
The de-Sitter solution (\ref{2.11}) corresponds to $H =H_0$.  From a physical point of view, such a solution has to be unstable in order to allow the transition of the universe to the radiation dominated era. To
show that this solution is unstable, consider a small deviation
from $H =H_0$:

\begin{eqnarray}
\label{2.14}
H=H_0(1+\delta)\,.
\end{eqnarray}
Substituting this in Eq. (\ref{2.13}) and linearizing in $\delta$ we obtain
\begin{eqnarray}
\label{2.15}
{\ddot \delta}+3H_0{\dot \delta}-M^2 \delta=0\,.
\end{eqnarray}
The two  solutions of (\ref{2.15}) are given
by $\delta=\exp(\alpha\,t)$ with

\begin{eqnarray}
\alpha=-\frac{3H_0}{2}\pm\sqrt{\frac{9H_0^2}{4}+M^2}\,.
\end{eqnarray}
The existence of a growing mode for  $\alpha>0$ indicates the
instability of the de-Sitter solution (\ref{2.11}). We have to stress  that the flat
space-time, $H =0$, is a stable solution of Eq. (\ref{2.13}). The linearization  in $H$ gives $2{\ddot H} =-M^2H$, which  has no growing
solutions for $M^2>0$. The linear approximation (\ref{2.15}) breaks down
when $\delta$ becomes $\sim1$.
The nonlinear evolution is achieved by  studying approximate solutions of Eq.
(\ref{2.13}) in various regimes.  We assume  that,   at the beginning, $H$ is near
$H_0$ and ${\dot H}$ is small, ${\dot H}<< H_0^2$. If $H>H_0$, then $H$ grows
without bounds. Such solutions are non-physical,  so one  has to take into account   the case
$H<H_0$. 

This situation is not  satisfactory since it implies a fine-tuning for initial conditions.  In some sense, this is a sort of {\it anthropic principle} where dynamics has to be selected {\it a priori}. Despite of this shortcoming, the problem can be addressed and solved in  the framework of Quantum Cosmology\cite{quantum} where unphysical initial conditions give rise to non-observable universes. The case $H>H_0$ falls into this set of conditions. 

With these considerations in mind, assuming $H(t)$ 
slowly varying, we have
\begin{eqnarray}
\label{2.16}
{\dot H}<<H^2\,, \quad {\ddot H}<<H{\dot H}\,.
\end{eqnarray}
The solution   of Eq. (\ref{2.13}) is
\begin{eqnarray}
\label{2.18}
H=H_0\tanh\left(\gamma-\frac{M^2t}{6H_0}\right)\,,
\end{eqnarray}
where $\gamma=\frac{1}{2}\ln\left(\frac{1}{\delta_0}\right)$ and $\delta_0$  is the magnitude of
$|H-H_0|/H_0$ for  $t =0$.
From Eq. (\ref{2.18}),  $H(t)$ changes on a time scale of the order
$\sim6H_0/M^2$. Sufficiently long inflation is obtained for  $M^2<< 6H_0^2$ .
The
solution (\ref{2.18}) is valid until the neglected terms become comparable to those we
kept in Eq.(\ref{2.13}). This happens for
 $H\sim M$. This means that during the  inflation, the expansion
rate gradually changes from  $H_0$ to $\sim M<<H_0$.
The scale factor $a (t)$ is found by integrating  Eq. (\ref{2.18}), that is
\begin{eqnarray}
a(t)=\frac{1}{H_0}\left[\frac{\cosh \gamma}{\cosh \gamma-\frac{M^2t}{6H_0}}\right]^{\frac{6H_0^2}{M^2}}\,.
\end{eqnarray}
For $t_\star-t \geq 6H_0/M^2$, this gives $a(t)=H_0 \exp(H_0\,t)$, and
for $t_\star<< 6H_0/M^2$, it is
\begin{eqnarray}
\label{2.19}
a(t)=\frac{1}{H_0}\left(\cosh \gamma\right)^{\frac{6H_0^2}{M^2}}\exp\left[-\frac{1}{12}M^2\left(t_\star-t\right)^2\right]\,.
\end{eqnarray}
The expansion rate  is
\begin{eqnarray}
\label{2.20}
H(t)=\frac{1}{6}\left(t_\star-t\right)\,.
\end{eqnarray}
The further evolution of the model can be achieved
for $H<<H_0$, when the term proportional to $H^4$ in Eq.
(\ref{2.13}) is
neglected, that is
\begin{eqnarray}
\label{2.21}
2H{\ddot H}+6H^2{\dot H}-{\dot H}^2+M^2H^2=0\,.
\end{eqnarray}
The {\it friction} term, $6H^2 {\dot H}$, is also small for $H<<M$.
An approximate solution of Eq. (\ref{2.21}) is
\begin{eqnarray}
\label{2.24}
H=\frac{4}{3t}\cos^2\left(\frac{M\,t}{2}\right)\left(1-\frac{\sin Mt}{Mt}\right)+{\cal O}\left(\frac{1}{t^3}\right).
\end{eqnarray}
Although $(Mt)^{-1}\sin Mt<<1$, this term has to be retained,
since its contribution to the derivatives of $H$ is not negligible.
 The scale factor is given by \cite{staro}
\begin{eqnarray}
\label{2.25}
a(t)=const \times t^{\frac{2}{3}}\left[1+\left(\frac{2}{3Mt}\right)\sin Mt+{\cal O}\left(\frac{1}{t^2}\right)\right]\,.
\end{eqnarray}
The expansion rate averaged over the oscillation period
 is
\begin{eqnarray}
\label{2.26}
{\bar H}=\frac{2}{3t}\,,
\end{eqnarray}
and  corresponds to the expansion law  ${\bar a}(t)\propto t^{2/3}$.  The oscillations of the expansion
rate in Eq. (\ref{2.24}) can be thought  as coherent
oscillations of a massive field describing scalar particles of
mass $M$ (the {\it scalarons}). The gravitational
effect of such {\it particles}  is similar to that of pressureless gas and leads to
the expansion law $ a\propto t^{2/3}$, that is to a matter dominated universe.

 However, after inflation, one expects a radiation-dominated epoch. The case of Eq. (\ref{2.26}) has been  obtained in the context of a homogeneous and isotropic universe without including  a radiation term. Since radiation evolves as $a^{-4}$, it will dominate over the effective dynamics due to $f(R)$ corrections driving the universe expansion after inflation. Thus, for the sake of completeness, one needs to include an additional $\propto a^{-4}$ term as soon as the Starobinky inflationary phase terminates\cite{vilenkin}.

\noindent The phenomenology of such a model is richer than that described here. One should consider also thermalization effects,  generation of gravitational waves, structure formation.  All these aspects are well discussed in literature \cite{vnb1,vnb2,vnb3,vnb4,vnb5}.
Here we want to stress  again that the shortcomings of early Standard Cosmological Model can be suitably solved by considering curvature terms generated by quantum effects in curved space. The paradigm is  to consider general classes of theories non-linear in the Ricci scalar (the so-called $f(R)$ gravity) and try to track the whole cosmic history up to dark energy epoch.

\section{The field equations of $f(R)$ gravity}
\label{quattro}

Having in mind the above results, we can now discuss a generic   $ f(R)$ function in the metric  formalism.  The geometric  approach adopted for the inflation (ultra-violet regime) may work even at current epoch (infra-red regime). Even if energy and size scales are completely different, an accelerating  behavior is recovered again by curvature corrections as it was first shown by Capozziello in 2002 \cite{rev5} and Carroll et al. in 2004 \cite{carroll}.
Let us consider the action \cite{PhysRepnostro}
\begin{equation}
 \label{act}
{\cal A}_{(curv)}= \int d^{4}x  \sqrt{-g} \, f(R) \,.
\end{equation}
The vanishing of the variation gives us the vacuum field equations:
\begin{eqnarray}\label{VAR12.32}
f'(R)R_{\mu\nu}-\frac{f(R)}{2} \, g_{\mu\nu} =\nabla_{\mu}
\nabla_{\nu}f'(R)-g_{\mu\nu}\Box
f'(R) \, ,
\end{eqnarray}
where the prime indicates the derivative with respect to the Ricci scalar $R$. The above equations  can be re-framed according to the Einstein-like form \cite{PhysRepnostro}

\begin{eqnarray}\label{VAR12.34}
G_{\mu\nu}=\frac{1}{f'(R)} \left\{
\nabla_{\mu}\nabla_{\nu} f'(R) - g_{\mu\nu}\Box f'(R)
+ g_{\mu\nu} \frac{ \left[ f(R)-f'(R) R \right]}{2}  \right\}\,.
\end{eqnarray}
The r.h.s. of Eq. (\ref{VAR12.34}) is thus reviewed as an effective energy-momentum  tensor. We name it as {\it
curvature   energy-momentum tensor} $T_{\mu\nu}^{(curv)}$. This tensor fuels the modified Einstein equations in terms of curvature corrections. Even though this interpretation is questionable, since the field equations depict a theory different from General Relativity, and one is forcing upon them the interpretation as effective Einstein equations, the scheme becomes therefore fruitful as it will be better clarified later. Further, considering the standard matter contribution, we get
 \begin{eqnarray}\label{h4}
G_{\mu\nu}&=&\frac{1}{f'(R)}\left\{\frac{1}{2}g_{\mu\nu}\left[f(R)-Rf'(R)\right]
+\nabla_{\mu}\nabla_{\nu} f'(R) -g_{\mu\nu}\Box f'(R)\right\}+ \frac{T^{(m)}_{\mu\nu}}{f'(R)}\nonumber\\&&=T^{(curv)}_{\mu\nu}+\frac{T^{(m)}_{\mu\nu}}{f'(R)}\,.
\end{eqnarray}
In the case of General Relativity,   $T^{(curv)}_{\mu\nu}$ becomes zero whereas the standard minimal coupling is easily reobtained for the matter
contribution. Thus, let us recast, for our convenience:
\begin{equation} \label{c6}
T^{(curv)}_{\mu\nu}\,=\,\frac{1}{f'(R)}\left\{\frac{1}{2}g_{\mu\nu}\left[f(R)-Rf'(R)\right]
+f'(R)^{;\mu\nu}(g_{\alpha\mu}g_{\beta\nu}-g_{\alpha\beta}g_{\mu\nu})
\right\}\,.
\end{equation}
Obviously this quantity  satisfies the Bianchi identities. Afterwards, our purpose is to demonstrate that it provides all the
requirements we need to tackle with the dark components of our cosmos. Depending on the precise scales, the curvature component may reproduce the dark energy \cite{mauro} and dark matter \cite{annalen,nav1,nav2,nav3} roles respectively. More precisely, even the coupling term $1/f'(R)$, entering the matter energy-momentum tensor, plays  a crucial role in the whole dynamics. This happens because it  affects all the physical processes ({\it e.g.} the nucleo-synthesis) and all the observable quantities (luminous, clustered, baryonic). In other words, the entire problem of understanding the universe dark components is naturally addressed, employing a self consistent theory where the interplay between geometry and  matter is reconsidered assuming non-linear contributions and non-minimal couplings in curvature invariants.

\section{Cosmology and curvature quintessence}
\label{cinque}

Reducing the action (\ref{act}) to a point-like, Friedmann-Robertson-Walker one, we
can write  the corresponding geometrical part as \cite{rev5}
\begin{equation}\label{8}
 {\cal A}_{(curv)}=\int dt {\cal L}(a, \dot{a}, R, \dot{R})\,{,}
\end{equation}
where, again, the dot indicates the derivative with respect to the cosmic time $t$. The scale factor $a=a(t)$ and the Ricci scalar $R$ may be assumed to be the  canonical variables. This appears as an arbitrary position since $R$ depends upon $a, \dot{a}, \ddot{a}$, but it is commonly employed in canonical
quantization procedures \cite{hamilton}.

The Ricci definition in terms of $a, \dot{a}, \ddot{a}$ involves a constraint that eliminates second and higher order derivatives in the action
(\ref{8}), providing a system of second order differential equations in terms of $\{a, R\}$. The action (\ref{8}) can be recast as
\begin{equation}\label{10}
{\cal A}_{(curv)}=2\pi^2\int dt \left\{ a^3f(R)-\lambda\left [ R+6\left (
\frac{\ddot{a}}{a}+\frac{\dot{a}^2}{a^2}+\frac{k}{a^2}\right)\right]\right\}\,{,}
\end{equation}
in which the Lagrange multiplier $\lambda$ has been obtained by varying with respect to the Ricci scalar $R$, giving
\begin{equation}\label{11}
\lambda=a^3f'(R)\,.
\end{equation}
It follows that the total point-like Lagrangian becomes
\begin{eqnarray}
\label{12}
{\cal L}&=&{\cal L}_{(curv)}+{\cal L}_{(m)} \nonumber\\&=& a^3\left[f(R)-R
f'(R)\right]+6a\dot{a}^2f'(R)+6a^2\dot{a}\dot{R}f''(R)-6k a
f'(R)+a^3p_{(m)}\,,\nonumber\\
\end{eqnarray}
which shows a canonical form in terms of the variables $\{a,\dot{a}, R. \dot{R} \}$. Here, the contribution of standard matter reduces to a pure pressure term. Hence, the Euler-Lagrange equations are
\begin{equation}
\label{13}
2\left(\frac{\ddot{a}}{a}\right)+\left(\frac{\dot{a}}{a}\right)^2+
\frac{k}{a^2}=-p_{(tot)},
\end{equation}
and
\begin{equation}
\label{14}
f''(R)\left\{R+6\left[\frac{\ddot{a}}{a}+\left(\frac{\dot{a}}{a}\right)^2+\frac{k}{a^2}\right]\right\}=0\,.
\end{equation}
In particular, Eq. (\ref{14}) is interpreted in terms of the Lagrange multiplier definition, guaranteeing the consistency of the approach.
Further, the dynamical system is completed by involving the following energy condition:
\begin{equation}
\label{15}
\left(\frac{\dot{a}}{a}\right)^2+\frac{k}{a^2}=\frac{1}{3}\rho_{(tot)}\,.
\end{equation}
In the above equations, we have
\begin{equation}
\label{17}
p_{(tot)}=p_{(curv)}+p_{(m)}\;\;\;\;\;\rho_{(tot)}=\rho_{(curv)}+\rho_{(m)}\,,
\end{equation}
in which we put in evidence both curvature and matter contributions to the whole cosmic fluid. We also inserted  the above non-minimal coupling factor $1/f'(R)$ into the matter term definition. From $T^{(curv)}_{\mu\nu}$, it is easy to get a curvature pressure definition
\begin{equation}
\label{18}
p_{(curv)}=\frac{1}{f'(R)}\left\{2\left(\frac{\dot{a}}{a}\right)\dot{R}f''(R)+\ddot{R}f''(R)+\dot{R}^2f'''(R)
-\frac{1}{2}\left[f(R)-Rf'(R)\right] \right\}\,,
\end{equation}
and a corresponding curvature density
\begin{equation}
\label{19}
\rho_{(curv)}=\frac{1}{f'(R)}\left\{\frac{1}{2}\left[f(R)-Rf'(R)\right]
-3\left(\frac{\dot{a}}{a}\right)\dot{R}f''(R) \right\}\,.
\end{equation}
Starting from the above formalism, the dark energy drawbacks and the phenomenon of the universe speed up can be described assuming this effective curvature term.
Combining Eq. (\ref{13}) and Eq. (\ref{15}), we get the Friedmann equation
\begin{equation}
\label{16}
\left(\frac{\ddot{a}}{a}\right)=-\frac{1}{6}\left[\rho_{(tot)}+3p_{(tot)}
\right]\,,
\end{equation}
in which it is evident that the acceleration depends upon the corresponding r.h.s. and then the acceleration is achieved for
\begin{equation}
\label{20} \rho_{(tot)}+ 3p_{(tot)}< 0\,,
\end{equation}
{\bf which means, from Eq. (\ref{16})}:
\begin{equation}
\label{21} \rho_{(curv)}\gg \rho_{m}\,.
\end{equation}
We assume  that the ordinary  matter components provide  non-negative pressure. Moreover, we assume that they are represented by standard  fluids, defined as $0\leq w_{(m)} \leq 1$.
Rephrasing it differently, viable conditions to observe cosmic acceleration depend on the relation
\begin{equation}
\label{22}
\rho_{(curv)}+3p_{(curv)}=\frac{3}{f'(R)}\left\{\dot{R}^2f'''(R)+\left(\frac{\dot{a}}{a}\right)\dot{R}f''(R)
+\ddot{R}f''(R)-\frac{1}{3}\left[f(R)-Rf'(R)\right]\right\}\,,
\end{equation}
which has to be compared with matter contribution which is not dominant, according to the observations. It has to be 
\begin{equation}
\label{23}
\frac{p_{(curv)}}{\rho_{(curv)}}=w_{(curv)}\,,
\qquad -1\leq w_{(curv)}<0\,.
\end{equation}
Particularly, the functional form of $f(R)$ represents the main ingredient to obtain curvature quintessence \cite{mauro,zigli1,zigli2,zigli3,zigli4}. Soon, it is clear that the simplest choice  to obtain the above prescriptions is to take into account a class of power-law
 solutions:
\begin{equation}
\label{24}
f(R)=f_0 R^n\,.
\end{equation}
 Inserting
Eqs. (\ref{24}) into the above dynamical system, we obtain, by Noether's symmetries \cite{rev5,basilakos}, the exact solutions
\begin{equation}
\label{25}  n=-1,\;\frac{3}{2}\,;\qquad \mbox{for}\quad k =0\,.
\end{equation}
In both the cases, the {\it deceleration} parameter is
\begin{equation}
\label{26}
q_0=-\frac{1}{2}\,,
\end{equation}
in perfect agreement with the expected values permitted in the case of cosmic acceleration. However, those solutions  cannot fit the whole cosmic history, together with some present shortcomings if confronted with data. However, they  can be considered as useful toy models to clarify how the problem of accelerating the universe can be addressed directly by $f(R)$ gravity \cite{raneg1,raneg2,raneg3,raneg4,raneg5}.

The case $n=3/2$ deserves a further discussion. Considering conformal transformation from Jordan frame to Einstein frame \cite{PhysRepnostro}, it is possible to give an explicit form for the scalar field potential that leads to the accelerated expansion. It is
\begin{equation}\label{27}
\tilde{g}_{\alpha\beta}\equiv
f'(R)g_{\alpha\beta}\,{,}\qquad
\varphi=\sqrt{\frac{3}{2}}\ln f'(R)\,{.}
\end{equation}
The conformal equivalence of the Lagrangians gives
\begin{equation}\label{28}
{\cal L}=\sqrt{-g}\,f_0R^{3/2}\longleftrightarrow
\tilde{\cal L}=\sqrt{-\tilde{g}}\left[-\frac{\tilde{R}}{2}+
\frac{1}{2}\nabla_{\mu}\varphi\nabla^{\mu}\varphi-V_0\exp\left(
\sqrt{\frac{2}{3}}\varphi\right)\right]\,{,}
\end{equation}
in our physical units.   This kind of model is particularly interesting to get inflation\cite{liddle}.  However, for the sake of completeness, it is relevant to notice that, in the Jordan frame,  Eq. (\ref{28}) may show problems with basic Solar System bounds. Further details may be found in literature\cite{tpab1,tpab2}.

For $n=3/2$,  and $k =0$, the general  solution of the system (\ref{13})-(\ref{15}) is\cite{rev5}

\begin{equation}\label{29}
a(t)=a_0\sqrt{c_4t^4+c_3t^3+c_2t^2+c_1t+c_0} \,{.}
\end{equation}
The integration constants $c_i$ are given by combining different initial conditions and their values definitively fix the cosmological evolution. For
example, if we consider $c_4\neq 0$, we obtain a power law inflation, whereas if the regime is dominated by the linear term in $c_1$, we find a
radiation-dominated epoch \cite{rubano}.

More realistic models can be worked out as reported in literature \cite{mauro,libroSV} but the general question is that the form of $f(R)$ function should be reconstructed by observational data. In next section, we will discuss in detail this problem.


\section{Cosmography}
\label{sei}

In this section, we introduce the basic demands of cosmography, giving particular emphasis to its standard usage to fix cosmographic
bounds on $f(R)$ and derivatives. In particular, to fix cosmological constraints on  curvature quintessence, it is important to find out a strategy which permits to reconstruct the universe expansion history at present time. Indeed, cosmography represents a method to constrain current time cosmology, without postulating any cosmological model \emph{a priori}. In so doing, dark energy's evolution can be directly framed in terms of cosmic data and $f(R)$ gravity can be featured by reconstructing numerical outcomes from the cosmographic coefficients. The corresponding \emph{cosmographic method} stands for a coarse grained technique to infer bounds on late time universe expansion history, rewriting quantities under interest in terms of cosmographic coefficients. Furthermore, cosmography is capable of discriminating  among competing  $f(R)$ models that are compatible with cosmographic predictions. From now on, we fix spatial curvature to be negligibly small, in order to get cosmography as a pure model independent treatment to bound the universe today \cite{planck,sound}. If scalar curvature is not fixed \emph{a priori}, a degeneracy problem occurs between the variation of acceleration and the spatial curvature density parameter $\Omega_k$ \cite{visse66}.

For our purposes, we simply use the $f(R)$ equation of state given by a geometrical fluid with curvature pressure $p_{curv}$ and we expand all quantities of interest into Taylor series around the present epoch, {\it i.e.} $z=0$. Typically, one may expand the Hubble parameter, the cosmological distances, the apparent magnitude modulus, the net pressure, and so forth \cite{orl,orl2,orl3,orl21,orl233,333,orl99}. All cosmographic coefficients are thus related to the derivatives of such Taylor expansions and can be bounded by cosmic data. In order to baptize such cosmographic coefficients and to permit one to handle cosmographic observables, it is possible to expand the scale factor $a(t)$ as
\begin{equation}\label{espansione}
a(t)=1+\sum_{n=1}^{\infty}\frac{d^na(t)}{dt^n}\Big|_{0}\Delta t^n\,,
\end{equation}
or more practically
\begin{eqnarray}\label{serie1a}
\frac{1-a(t)}{H_0} & \sim &   \Delta t - \frac{q_0}{2} H_0 \Delta t^2 +
\frac{j_0}{6} H_0^2 \Delta t^3 +   \frac{s_0}{24} H_0^3 \Delta t^4 +\ldots\,,
\end{eqnarray}
which displays the $a(t)$ Taylor series around $\Delta t\equiv t-t_0$, truncated at the fourth order. Usually, $q_0,j_0,s_0,\ldots$ are named the \emph{cosmographic series} (CS), representing scale factor derivatives evaluated at present time, i.e. at the redshift $z=0$. In particular, $q_0$ is the deceleration parameter that quantifies how much the universe accelerates today, $j_0$ is the jerk parameter and it is related to the variation of $q(t)$ in the past, whereas $s_0$ measures the change of $j(t)$ and it is commonly referred to as the snap parameter.

Usually, the today Hubble rate  $H_0$ enters the definition of the CS. However, since all coefficients may be  expressed in terms of $H(t)$, at all stages of the universe evolution, it would be better to consider $H_0$ as the parameter to set the CS \cite{mio}. In so doing, the cosmographic approach does not involve the definition of any cosmological model, becoming a powerful model independent method to fix $f(R)$ limits at late times. In other words, $H_0$ is a {\it prior} set from observational data. Therefore,  the CS is  defined as follows
\begin{eqnarray}\label{eq:CSoftime}
\frac{\dot{H}_0}{H_0^2}=-(1+q)\,, \quad \frac{\ddot{H}_0}{H_0^3}=j+3q+2\,, \quad \frac{H_0^{(3)}}{H_0^4}=s-4j-3q\left(q+4\right)-6\,.
\end{eqnarray}
All quantities are evaluated at present time $t=t_0$. In principle, the coefficients can be defined at all epochs by  considering  more general definitions as
\begin{equation} \label{eq:CScoeff}
     H(t) = \frac{1}{a}\frac{da}{dt}, \quad
    q(t) = -\frac{1}{a     H^2} \frac{d^2a}{dt^2}, \quad
    j(t) = \frac{1}{a     H^3} \frac{d^3a}{dt^3}\,\quad s(t)=\frac{1}{a     H^4} \frac{d^4a}{dt^4}\,.
\end{equation}
Thus, cosmography enables one to get a {\it snapshot} of the observable universe in terms of the CS, in order to reconstruct the universe cosmic evolution at different epochs \cite{turnercosmografia}. However, possible drawbacks are essentially based on the fact that current data are not accurate enough to fit significant intervals of convergence for  $z\gg1$. Moreover, there exist no  physical arguments to employ a particular cosmological distance than others, since all distances are physically well supported. In fact, all standard definitions  implicitly postulate that the universe is currently speeding up \cite{lix1,lix2}, since they are built up in terms of the photon distance $r_0$, i.e. the length that a photon travels from a light source at $r=r_0$ to a given reference point placed at $r=0$. The photon length definition leads to ${\displaystyle r_0 = \int_{t}^{t_0}{\frac{dt'}{a(t')}}}$ and depends on the scale factor only. Rephrasing those two problems differently, cosmographic expansions are plagued by a convergence problem due to truncated series, fitted with data in the interval $z\gg1$, and by a {\it duality problem}, since the correct cosmological distance to fit data is not known \emph{a priori}. To alleviate the convergence problem, an alternative   approach can provides the construction of  different redshift definitions, i.e. \emph{ad hoc} functions of the redshift $z$. Those re-parameterizing functions reduce the redshift intervals to tighter ranges \cite{gtm,gtm1,gtm2} and fulfill the conditions that all distance curves should not behave too steeply in the interval $z<1$. Once re-parameterized functions are built up to be one-to-one invertible, they can be directly compared with data \cite{mio}. Essentially, any viable re-parameterizations need, as basic requirements, to satisfy the following two properties:
\begin{subequations}\label{u}
\begin{align}
\mathcal{Z}&\rightarrow 1\quad z\rightarrow\infty\,,\\
\mathcal{Z}&\rightarrow 0\quad z\rightarrow0\,.
\end{align}
\end{subequations}
A simple example of reparametrization is offered by $\mathcal{Z}(z)\equiv\,z\,(1+z)^{-1}$. Afterwards, we list below three relevant definitions, in terms of $r_0$, as possible examples of cosmic distances:
\begin{subequations}\label{gb}
\begin{align}
    d_L  &=  a_0 r_0 (1+z) = r_0\,a(t)^{-1}\,,  \\
    d_F  &=  \frac{d_L}{(1+z)^{1/2}} = r_0\,a(t)^{-\frac{1}{2}}\,, \\
    d_A  &=  \frac{d_L}{(1+z)^2} = r_0\, a(t)\,,
\end{align}
\end{subequations}
respectively the luminosity, flux and angular distance. All the different cosmological distances assume the total number of photons is preserved \cite{bll} and all reduce at first order to
\begin{equation}\label{diconniente}
d_i\sim\frac{z}{H_0}\,,
\end{equation}
where $d_i$ represents the generic distance, i.e. $i=L;F;A$. Notice that all the above  distances can be rewritten in terms of auxiliary variables $\mathcal{Z}(z)$. Moreover, once $H_0$ is fixed, the series better converges, since its shape increases or decreases as $H_0$ decreases or increases respectively. Thus, fixing $H_0$ leads to determine a \emph{low redshift cosmographic setting value}, since all distances reduce to Eq. (\ref{diconniente}) at  first order of Taylor expansions.

This technique enables one to get viable cosmographic constraints that should be related somehow to $f(R)$ function and its derivatives. To do so, it is straightforward to  start from the definition of $R$ in terms of $H$, {\it i.e.} $R = -6 \left ( \dot{H} + 2 H^2\right )$ and by means of
\begin{equation}\label{redss}
\frac{d\log(1+z)}{dt}=-H(z)\,,
\end{equation}
having
\begin{equation}\label{eq: constr}
R = 6 \left[ (1+z)H\,H_{z} - 2 H^2\right]\,,
\end{equation}
where the subscript indicates the derivative with respect to the redshift $z$. We are assuming that the spatial curvature is $k=0$. Deriving $R$ at different orders allows to relate $R$ to $H$ and derivatives. Thus, since $H$ can be expanded as
\begin{equation}\label{kjhfkjhfjk}
H=H_0+\sum_{n=1}^{\infty}\frac{d^nH}{dz^n}z^n\,,
\end{equation}
it is easy to show
\begin{equation}\label{Hinz0}
\begin{split}
H_{z0}/H_0=\,& 1+q_0\,,\\
H_{2z0}/H_0=\,& j_0-q_0^2\,,\\
H_{3z0}/H_0=\,&-3j_0-4j_0q_0+q_0^2+3q_0^3-s_0\,,
\end{split}
\end{equation}
where we adopted the convention $H_{nz0}\equiv\frac{d^nH}{dz^n}\Big|_{0}$. Considering a pure matter term, evolving as dust, the cosmological Eqs. (\ref{13}) and (\ref{15}) can be recast, by a little algebra,  as
\begin{equation}\label{euna}
H^2 = \frac{1}{3} \left [ \rho_{(curv)} + \frac{\rho_{(m)}}{f'(R)} \right
]\,,
\end{equation}
\begin{equation}\label{edue}
2 \dot{H} + 3H^2= - p_{(curv)}\,.
\end{equation}
From the above expressions, deriving the $f(R)$ function, one gets
\begin{equation}
\begin{split}\label{zuzzu}
f'(R) =\, & R_z^{-1}f_z\,,\\
f''(R)=\, & (f_{2z}R_z - f_zR_{2z})R_z^{-3}\,,\\
f'''(R)=\, &\frac{f_{3z}}{R_z^3} - \frac{f_z\, R_{3z}+3f_{2z}\,
R_{2z}}{R_z^4}+\frac{3f_z\, R_{2z}^2}{R_z^5}\,,
\end{split}
\end{equation}
where we introduced the definition of $f(z)$. Indeed, since $R=R(z)$,  there exists a direct correspondence between $f(R)$ amd $f(z)$ functions: knowing $f(z)$ is equivalent to know $f(R)$ and viceversa. In cosmographic treatments, it is much easier to handle $f(z)$ than $f(R)$, due to the complexity of the modified Friedmann equations. Afterwards, since
\begin{equation}\label{hun44}
\dot{H}=-(1+z)HR_z\,,
\end{equation}
and
\begin{equation}\label{hun45}
\ddot{H}=(1+z)H\big[HR_z+(1+z)(H_zR_z+HR_{2z})\big]\,,
\end{equation}
we easily get
\begin{equation}\label{f0fz0fzz0dopo}
\begin{split}
\frac{f_0}{2H_0^2}=\,&-2+q_0\,,\\
\frac{f_{z0}}{6H_0^2} =\,&-2-q_0+j_0\,,\\
\frac{f_{2z0}}{6H_0^2}=\,&-2-4q_0-(2+q_0)j_0-s_0\,,
\end{split}
\end{equation}
which  to calculate $f(R)$ by a simple inverse procedure, once $f(z)$ and its derivatives are numerically known. Hence, determining numerical outcomes derived by cosmography, it is possible to bound $f(z)$ and its corresponding derivatives. In particular, if $f(z)$ is somehow constrained by cosmography, it naturally follows that $f(R)$ is bounded as well, since $R=R(z)$.

In particular, to obtain $f(R)$ and derivatives, one needs to know $R$ as a  function of the cosmographic parameters. Thus, we have
\begin{equation}\label{Rixxi}
\begin{split}
\frac{R_0}{6H_0^2}=\,&q_0-1\,,\\
\frac{R_{z0}}{6H_0^2} =\,&j_0-q0-2\,,\\
\frac{R_{2z0}}{6H_0^2}=\,&-\left(2+4q_0+2q_0^2+j_0(2+q_0)+s_0\right)\,.
\end{split}
\end{equation}

Moreover, it is possible to demonstrate that $f(z)$ and $f(R)$ decrease as the redshift increases. Analogously, the corresponding first derivatives negatively evolve as the redshift expands.

To get constraints on $f(z)$ and derivatives, one needs experimental procedures able to fix numerical outcomes on the CS. Two relevant data sets are for example given by the  Union 2.1 compilation \cite{kow} and the baryonic acoustic oscillation (BAO) \cite{sn32}. To fix viable constraints, it is possible to perform a Monte Carlo analysis based on minimizing the $\chi$ square functions. In the case of supernovae, we have as distance modulus,
\begin{equation}
\mu = 25 + 5 \log_{10} \frac{d_L}{Mpc}\,,
\end{equation}
and $\chi$ square function
\begin{equation}
\chi^{2}_{SN} =
\sum_{i}\frac{(\mu_{i}^{\mathrm{theor}}-\mu_{i}^{\mathrm{obs}})^{2}}
{\sigma_{i}^{2}}\,,
\end{equation}
while in case of BAO, we employ the measurable $\mathcal{A}$, defined  as
\begin{equation}
\mathcal{A}=\sqrt{\Omega_m}  \Big[\frac{H_0}{H(z_{BAO})}\Big]^{\frac{1}{3}}
\left[ \frac{1}{z_{BAO}}\int_0^{z_{BAO}}
\frac{H_0}{H(z)}dz\right]^{\frac{2}{3}}\,,
\end{equation}
with $z_{BAO}=0.35$ and the corresponding $\chi$ square:
\begin{equation}
\chi^{2}_{BAO}=\frac{1}{\nu}\left(\frac{\mathcal{A}-\mathcal{A}_{obs}}{\sigma_\mathcal{A}}\right)^2\,.
\end{equation}
Estimations of the CS may be performed through standard Bayesian techniques, i.e. maximizing the likelihood function:
$    \mathcal{L}_i
\propto \exp (-\chi_t^2/2 )
$,
where we define the total $\chi_t\equiv\chi_{SN}+\chi_{BAO}$. For the sake of completeness, one can expand the above cited causal distances $d_L$, $d_F$ and $d_A$, in terms of the redshift $z$, obtaining
\begin{eqnarray*}\label{dlinterminidiz}
d_L &=&  \frac{1}{H_0} \cdot \Bigl[ z + z^2 \cdot \Bigl(\frac{1}{2} - \frac{q_0}{2} \Bigr) +
    z^3 \cdot \Bigl(-\frac{1}{6} -\frac{j_0}{6} + \frac{q_0}{6} + \frac{q_0^2}{2} \Bigr)+ \nonumber\\
    &&+\, z^4 \cdot \Bigl( \frac{1}{12} + \frac{5 j_0}{24} - \frac{q_0}{12} + \frac{5 j_0 q_0}{12} -
    \frac{5 q_0^2}{8} - \frac{5 q_0^3}{8} + \frac{s_0}{24} \Bigr)+\ldots\Bigr]\,,
\end{eqnarray*}

\begin{eqnarray*}
    d_F & = & \frac{1}{H_0} \cdot \Bigl[ z - z^2 \cdot \frac{q_0}{2} + z^3 \cdot \Bigl(-\frac{1}{24}
    -\frac{j_0}{6} + \frac{5q_0}{12} + \frac{q_0^2}{2} \Bigr)+ \nonumber\\
    &&+\, z^4 \cdot \Bigl( \frac{1}{24} + \frac{7 j_0}{24} - \frac{17 q_0}{48} + \frac{5 j_0 q_0}{12}
    - \frac{7 q_0^2}{8} - \frac{5 q_0^3}{8} + \frac{s_0}{24} \Bigr)+\ldots\Bigr]\,,
\end{eqnarray*}
and
\begin{eqnarray*}
    d_A & = & \frac{1}{H_0} \cdot \Bigl[ z + z^2 \cdot \Bigl( -\frac{3}{2} - \frac{q_0}{2} \Bigr)
    + z^3 \cdot \Bigl( \frac{11}{6} -\frac{j_0}{6} + \frac{7 q_0}{6} + \frac{q_0^2}{2} \Bigr) + \nonumber\\
    &&+\, z^4 \cdot \Bigl( -\frac{25}{12} + \frac{13 j_0}{24} - \frac{23 q_0}{12} + \frac{5 j_0 q_0}{12}
    - \frac{13 q_0^2}{8} - \frac{5 q_0^3}{8} + \frac{s_0}{24} \Bigr)+\ldots\Bigr]\,.
\end{eqnarray*}
Those expansions enter the definitions of $\chi_t^2$ and, after a numerical procedure, it is easy to bound the CS. Afterwards, keeping in mind $q_0,j_0,s_0$, and using Eqs. (\ref{f0fz0fzz0dopo}), it is possible to numerically fix $f(z)$ and derivatives. Analogously, from Eqs. (\ref{zuzzu}), it is possible to constrain $f(R)$. This technique enables the determination of $f(R)$ as the universe expands and consequently our numerical outcomes lead to frame the dark energy effects in terms of a pure curvature fluid. Cosmographic indications suggest that the cosmological standard model is extended by means of a logarithmic correction, as follows\cite{faro}
\begin{equation}\label{h}
{H}(z)={H}_0\sqrt{\Omega_m(1+z)^3+\log(\alpha+\beta z)}\,,
\end{equation}
where $\beta$ is a free constant, whereas $\alpha=\exp(1-\Omega_m)$. In other words, Eq. (\ref{h}) represents an \emph{effective Hubble rate},  numerically reconstructed, by employing $f(R)$ corrections, set through cosmographic results.

\noindent This prescription provides an approximate form of $f(z)$ given by
\begin{equation}\label{jhd}
f(z)= \tilde f_0+\frac{1}{1+z}+\tilde f_1 (1+z)^{\sigma_1}+\tilde f_2(1+z)^{\sigma_2}\,,
\end{equation}
which well adapts its shape to data, with negligible departures from $z\ll1$ to $z\sim2$, for the intervals $\tilde f_0\sim-10$, $\tilde f_1\sim 7$, $\tilde f_2\sim-3.7$, $\sigma_1=1$ and $\sigma_2=2$. Those results are consistent with the cosmographic ranges of $f_0$ and $f_{z0}$. As example of direct fittings of $f(z)$ and derivatives, we report some experimental results in Tab. I, while in Tab. II we report the corresponding cosmographic parameters\cite{noi1}.

\begin{table*}
\caption{{\footnotesize Best fits of the parameters $H_0$, $f_0$, $f_{z0}$ and $f_{2z0}$ for three statistical models, i.e. A, B and C, corresponding to three different orders of the cosmographic Taylor expansion, respectively the second, third and fourth orders.  }}
\begin{tabular}{c|c|c|c} 
\hline\hline\hline 
$\qquad${\small Parameter}$\qquad$   &  $\qquad$ Model A $\qquad$& $\qquad$ Model B $\qquad$ & $\qquad$ Model C $\qquad$ \\

\hline
$H_0$                 & {\small $77.23$}{\tiny${}_{-1.82}^{+0.84}$}       & {\small $75.69$}{\tiny${}_{-1.99}^{+2.03}$}       & {\small $71.30$}{\tiny${}_{-1.91}^{+1.92}$}     \\[0.8ex]
\hline
$10^{-4} f_0$         & {\small $-3.324$}{\tiny${}_{-0.230}^{+0.227}$}    & {\small $-3.144$}{\tiny${}_{-0.332}^{+0.320}$}    & {\small $-2.669$}{\tiny${}_{-0.284}^{+0.287}$}  \\[0.8ex]
\hline
$10^{-4} f_{z0}$      & {\small $3.636$}{\tiny${}_{-1.735}^{+1.751}$}     & {\small $-1.510$}{\tiny${}_{-5.656}^{+5.694}$}    & {\small $-1.794$}{\tiny${}_{-4.200}^{+4.834}$}  \\[0.8ex]
\hline
$10^{-5} f_{2z0}$     & {\small $-2.202$}{\tiny${}_{-0.973}^{+0.965}$}    & {\small $2.276$}{\tiny${}_{-2.032}^{+2.339}$}     & {\small $0.499$}{\tiny${}_{-2.049}^{+2.192}$}   \\[0.8ex]
\hline

\hline\hline\hline
\end{tabular}

{\footnotesize
Table I.
Best fits of the parameters $H_0$, $f_0$, $f_{z0}$ and $f_{2z0}$ for three statistical models, i.e. A, B and C, corresponding to three different orders of the cosmographic Taylor expansion, respectively the second, third and fourth orders. $H_0$ is given in Km/s/Mpc.}
\label{table:summary} 
\end{table*}

\begin{table*}
\caption{{\footnotesize Values of the cosmographic coefficients for three statistical models, i.e. A, B and C, corresponding to three different orders of the cosmographic Taylor expansion, respectively the second, third and fourth orders.}}
\begin{tabular}{c|c|c|c}
\hline\hline\hline
$\qquad${\small Parameter}$\qquad$   &  $\qquad$ Model A $\qquad$& $\qquad$ Model B $\qquad$ & $\qquad$ Model C $\qquad$ \\

\hline
$q_0$         & {$-0.786$}{\tiny${}_{-0.324}^{+0.251}$}    & {$-0.744$}{\tiny${}_{-0.434}^{+0.426}$}    & {$-0.625$}{\tiny${}_{-0.420}^{+0.424}$}  \\[2.3ex]
\hline
$j_0$      & {$2.229$}{\tiny${}_{-0.761}^{+0.718}$}      & {$0.817$}{\tiny${}_{-2.102}^{+2.106}$}    & {$0.787$}{\tiny${}_{-1.83}^{+2.04}$} \\[2.3ex]
\hline
$s_0$     & {$-7.713$}{\tiny${}_{-5.372}^{+4.997}$}      & {$-6.671$}{\tiny${}_{-10.295}^{+11.15}$}    & {$-2.217$}{\tiny${}_{-11.15}^{+11.93}$}  \\[2.3ex]
\hline

\hline\hline\hline
\end{tabular}

{\footnotesize
Table II. Table of numerical results for the CS; the numerical values are given at $z=0$, corresponding to three different orders of the cosmographic Taylor expansion, respectively the second, third and fourth orders.}
\label{table:CS} 
\end{table*}
All numerics provide a slightly lower dark energy pressure, with small corrections to a constant dark energy term. This fact suggests that the curvature dark energy is not described by a pure cosmological constant. The curvature quintessence seems to evolve in agreement with the following limits
\begin{equation}
\label{summary}
f_{0}<0\,,\quad
f_{z0}>0\,,\quad
f_{2z0}<0\,,\quad
p_{curv}<p_{\Lambda CDM}\,,
\end{equation}
where $p_{\Lambda CDM}$ is the pressure of the standard cosmological model. Moreover, it is a matter of fact that the absolute values of each variable increases as one performs fits using $d_F$ and $d_A$, giving reasonable departures from the cosmological standard model. Determining the limits on $f(z)$ and derivatives,  it is easy to show the corresponding bounds on $f(R)$, listed in Tab. III.

\begin{table*}
\caption{{\footnotesize Values of $f(R)$ and its derivatives for three statistical models, i.e. A, B and C, corresponding to three different orders of the cosmographic Taylor expansion, respectively the second, third and fourth orders.}}
\begin{tabular}{c|c|c|c} 
\hline\hline\hline 
$\qquad${\small Parameter}$\qquad$   &  $\qquad$ Model A $\qquad$& $\qquad$ Model B $\qquad$ & $\qquad$ Model C $\qquad$ \\

\hline
$f(R_0)$         & {$-3.324$}{\tiny${}_{-0.230}^{+0.227}$}    & {$-3.144$}{\tiny${}_{-0.332}^{+0.320}$}    & {$-2.669$}{\tiny${}_{-0.284}^{+0.287}$}  \\[2.3ex]
\hline
$f'(R_0)$      & {$1$}{\tiny${}_{-2.7\cdot10^{-16}}^{+2.6\cdot10^{-16}}$}     & {$1$}{\tiny${}_{-1.8\cdot10^{-15}}^{+1.8\cdot10^{-15}}$}    & {$1$}{\tiny${}_{-5.3\cdot10^{-16}}^{+5.8\cdot10^{-16}}$}  \\[2.3ex]
\hline
$f^{''}(R_0)$     & {$5.9\cdot10^{-20}$}{\tiny${}_{-7.8\cdot10^{-20}}^{+7.3\cdot10^{-20} }$}     & {$-4.1\cdot10^{-19}$}{\tiny${}_{-4.7\cdot10^{-18}}^{+4.8\cdot10^{-18} }$}     & {$-1.2\cdot10^{-19}$}{\tiny${}_{-6.8\cdot10^{-19}}^{+7.8\cdot10^{-19} }$}   \\[2.3ex]
\hline

\hline\hline\hline
\end{tabular}

{\footnotesize
Table III. Values of $f(R)$ and its derivatives for three statistical models, i.e. A, B and C, corresponding to three different orders of the cosmographic Taylor expansion, respectively the second, third and fourth orders.}
\label{table:summary2} 
\end{table*}

\begin{table*}
\caption{{\footnotesize Values of $f(R)$ and its derivatives for three statistical models, i.e. A, B and C, corresponding to three different orders of the cosmographic Taylor expansion, respectively the second, third and fourth orders.}}
\begin{tabular}{c|c|c|c} 
\hline\hline\hline 
$\qquad${\small Parameter}$\qquad$   &  $\qquad$ Model A $\qquad$& $\qquad$ Model B $\qquad$ & $\qquad$ Model C $\qquad$ \\

\hline
$R_0$         & {$-10.716$}{\tiny${}_{-1.944}^{+1.506}$}    & {$-10.464$}{\tiny${}_{-2.604}^{+2.556}$}    & {$-9.750$}{\tiny${}_{-2.520}^{+2.544}$}  \\[2.3ex]
\hline
$R'_0$      & {$6.090$}{\tiny${}_{-2.622}^{+2.802}$}     & {$-2.634$}{\tiny${}_{-10.008}^{+10.079}$}    & {$-3.528$}{\tiny${}_{-8.460}^{+9.696}$}  \\[2.3ex]
\hline
$R^{''}_0$     & {$29.492$}{\tiny${}_{-41.695}^{+51.033}$}     & {$33.082$}{\tiny${}_{-95.037}^{+74.670 }$}     & {$5.122$}{\tiny${}_{-101.575}^{+81.032}$}   \\[2.3ex]
\hline

\hline\hline\hline
\end{tabular}

{\footnotesize
Table IV. Values of $R$ and its derivatives for three statistical models, i.e. A, B and C, corresponding to three different orders of the cosmographic Taylor expansion, respectively the second, third and fourth orders. The numerical results have been obtained, in power of $H_0^2$, using the numerical outcomes inferred from Tab. II and the expressions of Eqs. (\ref{Rixxi}). }
\label{table:summary3} 
\end{table*}
From Tabs. I, II and III, we are able to describe the $f(z)$, $f(R)$ functions at late times, fixing the corresponding numerical bounds which represent the cosmographic settings (Tab. I) on $f(z)$ and derivatives. In other words, we are able to depict the universe expansion history in the observable limit of small redshift.  In Tab. IV, we summarize the numerical outcomes inferred for $R$ and derivatives using the cosmographic results.

To extrapolate the behavior of the $f(R)$ function at different stages of the universe evolution, one can use Eq. (\ref{zuzzu}), Eqs. (\ref{h}),  and (\ref{jhd})  to perform an inverse procedure and define a corresponding $f(R)$ function at our time. In fact, since
\begin{equation}\label{reconstrFR}
f(R)=\int \frac{df}{dz}(R_z)^{-1}dz + K_{cs}\,,
\end{equation}
with $K_{cs}$ a cosmographic integration constant.   Let us notice that $K_{cs}$ is not related to the cosmological constant. Instead, the constant $K_{cs}$ estimates the numerical difference between putting by hand $z$ in function of $R$ within $f(z)$ and evaluating the integral ${\displaystyle \int \frac{df}{dz}(R_z)^{-1}dz}$.
Both the procedures enable one to extrapolate the numerical $f(R)$ function, knowing $f(z)$, although they differ from a constant, {\it i.e.} $K_{cs}$.
Indeed, integrating  the $f(z)$ function over the whole redshifts may increase or decrease the final shape of the numerical $f(R)$ function with respect to directly substitute $z$ in function of $R$ into  $f(z)$.
The increasing or decreasing factor is approximatively a constant, $K_{cs}$, which is due to the  integration performed in Eq. (\ref{reconstrFR}).

It is also possible to determine the $f(R)$ shape, by keeping in mind the form of $f(z)$, for redshift intervals $z>1$. A significative  extension to higher redshifts employing Eq. (\ref{h}) is possible by numerically solving the modified Friedmann equations.  The above  procedure allows to reproduce the effective dynamics of dark energy without postulating the existence of a cosmological constant. Below, we report a reasonable cosmographic reconstruction that has been achieved by the following $f(R)$ reconstruction: \cite{noi1,noi}
\begin{equation}\label{fnostra}
\begin{split}
f(R)=&\frac{1}{2(a+b+c)e\pi R_0^2}\bigg\{
\Lambda R_0^2\bigg[\,
2a\pi e^{R/R_0}\\
&+e\bigg(6b+(a+2c)\pi+8b\, \arctan\left(\frac{R}{R_0}\right)\bigg)
\bigg]\\
&+e\mathcal{R}\bigg[
2R_0\big((a+b+c)\pi R_0-4b\Lambda\big)\\
&+(2b-a\pi)\Lambda R
\bigg]-2ce\pi \Lambda(R-R_0)^2\sin\left(\frac{2\pi R }{R_0}\right)
\bigg\}\,,
\end{split}
\end{equation}
where $a,b,c$ represent three free parameters of the model related to integration constants. The model has been obtained by considering the cosmographic results on CS as initial settings, and fitting the numerical outcomes obtained from Eqs. (\ref{euna}), (\ref{edue}). The corresponding $f(R)$ function  passes several  experimental bounds at small redshifts \cite{aggiuntaL} with high agreement at higher redshifts. Comparing cosmographic results with the approximate function given in Eq. (\ref{fnostra}), we find
\begin{equation}\label{constrainta}
a\sim145.5_{-8.73}^{+11.64}\,,\qquad
b\sim-148_{-8.88}^{+11.44}\,,\qquad
c\sim1_{-0.06}^{+0.08}\,,
\end{equation}
which represent the numerical outcomes of the free parameters involved in Eq. (\ref{fnostra}).  Typically the errors are estimated not to exceed the limit of $6\div 8\%$. Thus for each coefficients, the errors bars $\delta a, \delta b, \delta c$ have been reported into the intervals $\delta a\sim[8.73,11.64]\,,
\delta b\sim[8.88,11.84]\,,
\delta c\sim[0.06,0.08]\,$ respectively.

Other typologies of viable candidates can be determined shifting the cosmographic outcomes and consequently changing the initial settings due to cosmography. The scheme proposed here is able to describe the universe dynamics at small redshifts, by means of $f(R)$ cosmography. This procedure matches  the early phases  of $f(R)$ cosmology  with present time  and it is of great importance in order to reconstruct robust forms of $f(R)$ function.  Future developments should allow to  refine the  $f(R)$ paradigm by relating  late time results with  high redshift data.

\section{Outlooks and perspectives}
\label{sette}
In this paper, we have outlined some of the main features of $f(R)$ cosmology, with no claim to completeness. Our aim has been to show that such an alternative approach to cosmology directly derives from a natural extension of General Relativity and can be based on fundamental physics since comes out from quantum field theory formulated on curved space. In principle, $f(R)$ gravity could trace back from late type cosmology up to inflation, if reliable  models are suitably constructed by matching observational data at various redshift regimes. Cosmographic analysis greatly aids in this task as soon as cosmographic parameters are
  derived from $f(R)$ functions and their derivatives, from one side,  and self-consistently are matched with data, on the other side. In principle, this procedure  could be extended up to inflation\cite{bicep} addressing also the large scale structure\cite{wu}, but the big challenge is to find out reliable and  homogeneous datasets which allow realistic fittings at any cosmic epochs.

From a genuine theoretical point of view, there is no final $f(R)$, or alternative gravity model, today capable of addressing all the cosmological dynamics, however, despite of this lack, the approach seems very promising to encompass both problems of inflation and dark side. The forthcoming space experiments like EUCLID\cite{euclid} could realistically support or rule out definitely this view.

\section*{Acknowledgments}

SC and MDL  acknowledge INFN Sez. di Napoli (Iniziative Specifiche CQSKY and TEONGRAV) for financial support. OL wants to express his gratitude to Hernando Quevedo for discussions. OL is supported by the European PONa3 00038F1 KM3NET (INFN) Project.


\end{document}